**Linear rather than exponential decay: a mathematical model and underlying mechanisms**

Irina Kareva[1*], Georgy Karev[2]

[1*] Computational and Modeling Sciences Center, Arizona State University, Tempe, AZ, 85287, USA. Email: ikareva@asu.edu

[2] National Center for Biotechnology Information, National Institutes of Health - Bldg. 38A, 8600 Rockville Pike, Bethesda, MD 20894, USA. Email: karev@ncbi.nlm.nih.gov

**Abstract**

Some populations, such as red blood cells (RBCs), exhibit a pattern of population decline that is closer to linear rather than exponential, which has proven to be unexpectedly challenging to describe with a single simple mathematical model. Here we show that a sub-exponential model of population extinction can approximate very well the experimental curves of RBC extinction, and that one possible mechanism underlying sub-exponential decay is population heterogeneity with respect to death rates of individuals. We further show that a sub-exponential model of population decline can be derived within the frameworks of frequency-dependent model of population extinction if there exists heterogeneity with respect to mortality rates such that their initial distribution is the Gamma distribution. Notably, specific biological mechanisms that may result in linear pattern of population decay may be different depending on the specific biological system; however, in the end they must converge to individual death rates being different within the population, since uniform death rates would result in exponential population decline. As such, the proposed model is not intended to describe the complex dynamics of RBC biology but instead can provide a way to phenomenologically describe linear decay of population size. We briefly discuss the potential utility of this model for describing effects of drugs that may cause RBC depletion and conclude with a suggestion that this tool can provide a lens for discovering linear extinction patterns in other populations, which may have previously been overlooked.

**Introduction**

Mathematical models of population extinction are typically described by exponential decay, where decrease in population size is proportional to the number of individuals currently present in the population. Such models take the form of $N(t)' = -kN(t)$, where $N(t)$ is the population size that changes over time, and *k* is the per capita death rate. However, such models do not always accurately describe experimental data, such as that of decay of erythrocytes, or red blood cells (RBCs). Instead, RBC elimination appears to be linear, as has been shown experimentally in for instance, Mock et al. (Mock et al. 2011). In this study, the authors labeled aliquots of autologous RBCs from healthy volunteers and tracked proportion of circulating labeled cells by flow cytometry over 20 weeks. RBC survival in subjects followed a pattern of extinction that is captured more closely by linear function rather than exponential (Fig 2 and 3 in (Mock et al. 2011)), with exponential decline observed only in the last several days.

A natural approach to describing linear death may appear to involve introduction of constant death rate, i.e., $N(t) = -k$, which may fit the data in early stages, but which soon predicts negative population size unless a mechanistic switch is introduced numerically at $N(t) = 0$. Mathematical representation of linear death rate using a single model presents a surprising challenge (Quinlivan 2008), which has been addressed with PDEs (Bélair et al. 1995; Higgins & Mahadevan 2010) or integro-differential equations (Shrestha et al. 2016). However, a single ODE model to successfully describe linear population decay to our knowledge does not yet exist, a gap that we aim to fill here.

Here, we show that a sub-exponential ODE model of population extinction can capture linear pattern of population decline. We also show that dynamics of such a model stem from intrinsic heterogeneity of the underlying population with respect to death rates. We use the model to fit experimental data on RBC extinction obtained from the literature and propose a possible interpretation with respect to RBC biology, a hypothesis that stems from the model itself (subscribing to the approach of observation leading to theory that can then lead to experimental validation). We conclude with a brief discussion of the potential for linear death patterns to be more ubiquitous than previously assumed.

**Model description**

Consider a model of population extinction given by the following equation:

$$\frac{dN}{dt} = -kN^p, \quad (1)$$

where instead of $p=1$ (typical exponential death), $0 < p < 1$.

As one can see in Figure 1, such a model can indeed predict linear extinction rate depending on initial population size $N(0)$, as well as different values of parameters $p<1$ and $k$. Notably, the larger the value of $p$ or the lower the value of $k$, the closer the curve is to describing exponential decay; the smaller the value of $p>0$, the closer the curve is to describing linear decay.

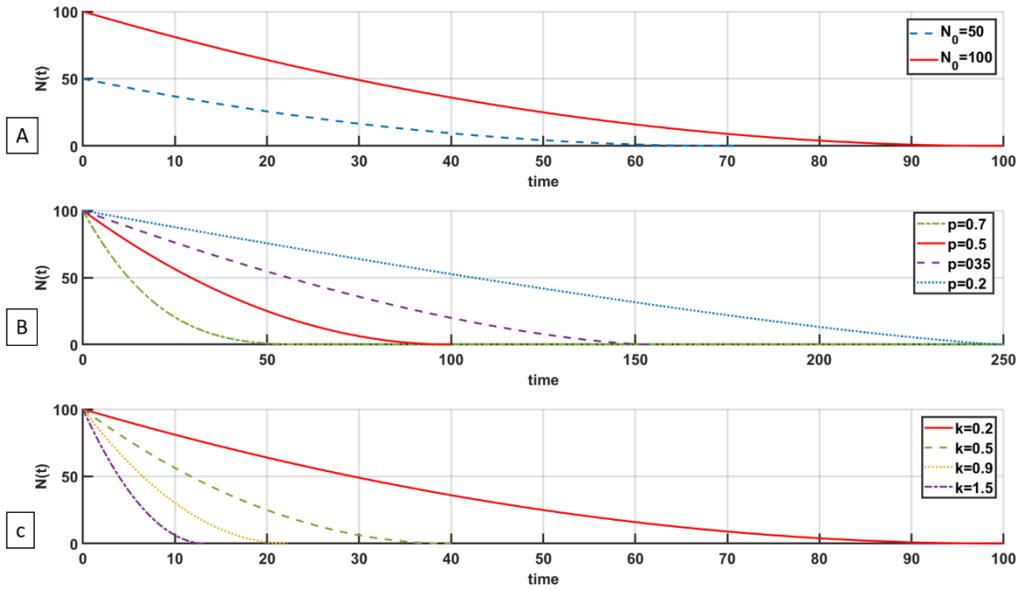

**Figure 1**. Dynamics predicted by Equation (1). As one can see, the shape of the curve can depend on (A) variations in $N_0$; $p=0.5$, $k=0.2$, (B) variations in $p$; $N_0=100$, $k=0.2$, and (C) variations in $k$; $N_0=100$, $p=0.5$. For reference, red solid line in all three panels has parameters $N_0=100$, $p=0.5$, $k=0.2$.

Solution to Equation (1) exists only for $t < \dfrac{N(0)^{1-p}}{k(1-p)}$ and is given by

$$N(t) = N(0)(1 - N(0)^{p-1} k(1-p)t)_+^{\frac{1}{1-p}}. \quad (2)$$

Here and henceforth the subscript "+" means a positive part of corresponding expression.

According to Equation (2), the population has a finite life span; it becomes completely extinct at the moment $T$, i.e. $N(T_{ext}) = 0$, where extinction moment $T_{ext}$ is given by

$$T_{ext} = \frac{N(0)^{1-p}}{k(1-p)} \qquad (3)$$

Full details of derivation and more complete analysis of these types of models can be found in (Karev & Kareva 2016), as well as in Chapter 10 of (Kareva & Karev 2019). For the purposes of this discussion, it is important to emphasize that time to extinction is explicitly defined by the three parameters of the model: initial population size $N(0)$, $p < 1$ and $k$. Dependence of $T_{ext}$ on parameters $p < 1$ and $k$ is shown in Figure 2.

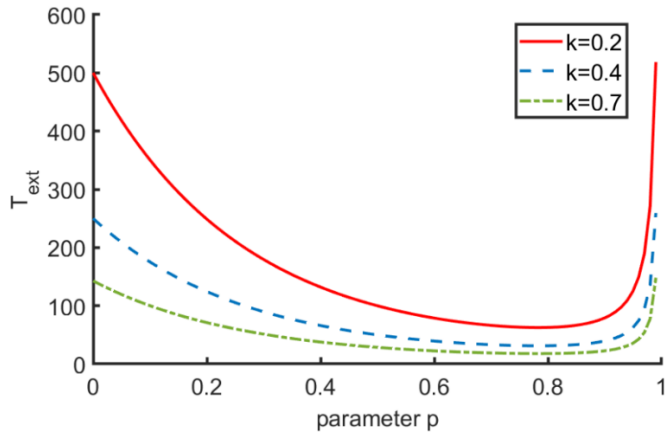

**Figure 2**. Dependence of time to extinction on parameters $p$ and $k$. Note that as $p \to 1$, $T_{ext} \to \infty$, as is expected for exponential decay model.

Now, using $T_{ext}$ as defined in Equation (3), we can rewrite the explicit solution for calculating total population size $N(t)$ given by Equation (2) as

$$N(t) = N(0)(1 - t/T)_+^{1/(1-p)} \qquad (4)$$

Setting $T_{ext} = 140$ (the mean lifespan of RBCs (Arias & Arias 2017)), we can now fit experimentally observed RBC extinction data using either explicit solution in Equation (4), or the original differential equation (1).

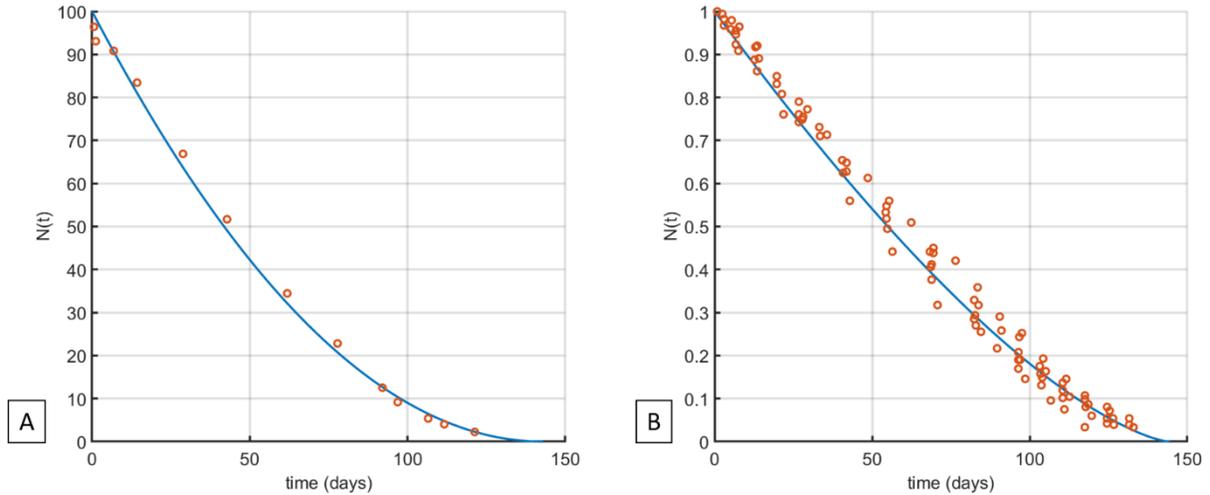

**Figure 3**. Fits of experimentally reported data using Equation (1). Blue lines represent model fits, while digitized data appoints are shown with red circles. (A) Fit of data set digitized from Franco, 2009 (Franco 2009). Parameters are $N_0 = 100$, $k = 0.14$, $p = 0.5$. (B) Fit of data set digitized from Shrestha et al. (Shrestha et al. 2016). Parameters are $N_0 = 1$, $k = 0.01$, $p = 0.3$. Fit was obtained using *fmincon* function in Matlab 2019a.

Note that population decay is linear until the final time points, where it becomes exponential (notably, larger values of parameter $p$ provide a smoother fit at later time points, as is expected, since with $p \to 1$ the model approaches exponential decay). Such linear-exponential dynamics are a characteristic of inhomogeneous frequency-dependent models, which will be discussed next.

*Underlying mechanisms of linear decay described by the sub-exponential model*

One way to infer a possible mechanism underlying linear extinction is as follows. The zero-order linear extinction $N(t)' = -k$ can be rewritten as

$$N(t)' = -kN(t)\frac{1}{N(t)}. \qquad (5)$$

If each individual in the population has its own death rate, then the population can be subdivided into individuals, or clones $l(t,a)$, where each individual is characterized by their own value of parameter $a$; total population size is then given by $N(t) = \sum_a l(t,a)$. Rewriting Equation (5) in terms of $l(t,a)$ leads to the following equation:

$$\frac{dl(t,a)}{dt} = -\frac{kal(t,a)}{N(t)} = -kaP(t,a) \tag{6}$$

It was proven in Theorem 2 in (Karev & Kareva 2016) that Equations (1) and (2) describe a population described by a frequency-dependent model (6) if individual death rate parameter $a$ comes from initial Gamma distribution, whose pdf is given by

$$P(0,a) = \frac{a^{\rho-1} e^{-\frac{a}{\beta}}}{\beta^\rho \Gamma(\rho)}, \ a > 0, \rho = \frac{1}{p}, \beta = pN(0)^p. \tag{7}$$

It was also proven that the current distribution of the death rate parameter is also Gamma-distributed with parameters

$$\rho = \frac{1}{p}, \ \beta = pN(t)^p \tag{8}$$

and

$$N(t) = \int_0^\infty l(t,a) P(t,a) da = N(0)(1 + kN(0)^{-1+p}(p-1)t)_+^{\frac{1}{1-p}} \tag{9}$$

The current mean value of the death rate is

$$E^t[a] = \rho\beta(t) = N(t)^p. \tag{10}$$

Hence, $E^t[a] \to 0$ as $t \to T_{ext}$. These results are derived in (Karev & Kareva 2016).

Notably, truncated Gamma distribution as the initial distribution is more realistic than full Gamma distribution because the latter allows arbitrary large values of the death rate. On the other hand, clones with large values of the death rate will be quickly eliminated from the population so one can realistically assume that the difference in these two cases can be negligible. Note that inhomogeneous model (6) with truncated Gamma-distribution does not allow a simple analytical solution similar to (2), but nevertheless can be solved with the help of the method used in (Karev & Kareva 2016).

These results can be summarized as follows: if population extinction dynamics are not exponential but linear, they can be described by sub-exponential extinction model as given by Equation (1), which in turn describes the dynamics of a heterogeneous population, where individuals with individual death rate parameters die proportionally to their frequency in the population. This can provide a novel model-driven hypothesis for mechanisms underlying biology of populations with linear decay, which will be addressed in the discussion.

*Modeling RBC turnover*

The model introduced in Equation (1) can now be used to create a simple model of cell turnover. The simplest approach is to introduce a constant inflow rate $a$, such that a homeostatic level of cells is maintained over time in the absence of intervention:

$$N(t)' = a - kN(t)^p, \qquad (11)$$

with $p < 1$.

This model can now be used to simulate the effect on cell population of drugs that can cause cell death. One way to capture this is to introduce an additional death term into Equation (11) that describes additional cell mortality as a function of concentration of hypothetical drug $C(t)$:

$$N(t)' = \underbrace{a - kN(t)^p}_{\text{normal turnover}} - \underbrace{k_1 C(t) N(t)}_{\text{death by drug}}, \qquad (12)$$
$$C(t)' = -k_e C(t)$$

where $C(t)$ is change in drug concentration over time, $k_e$ is drug clearance rate and $k_1$ is rate of cell kill that is proportional to drug concentration.

A simulation of the impact of such simulated drug can be seen in Figure 4. When a more complex and physiologically accurate model of RBC dynamics is not necessary for the purposes of the scientific question, this simple phenomenological model can be used to describe drug-induced anemia in pre-clinical simulations.

Note that all parameters pertaining to drug concentration profile provided in Figure 4, as well as the effect of such drug on the cells, were chosen arbitrarily for illustration purposes. The model, however, does capture key features of these interactions and can be easily modified to fit characteristics of specific drugs.

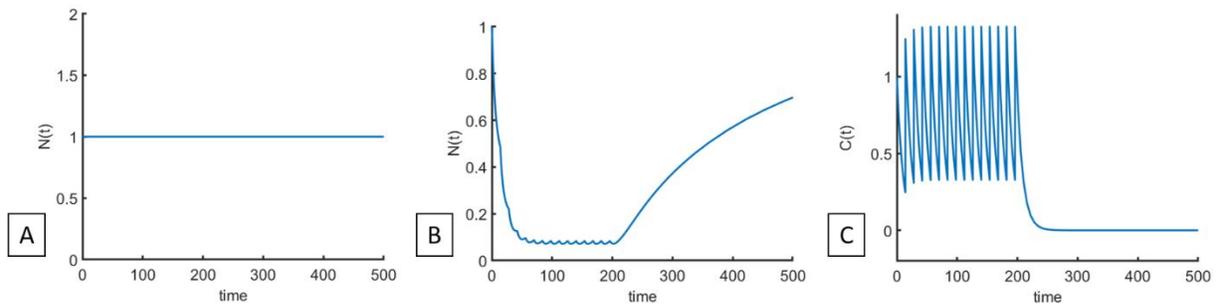

**Figure 4**. Cell turnover model with linear decay, with and without simulated treatment. (A) Normal cell homeostasis model as described by Equation (11). Model parameters were taken as $a = 0.01$, $N_0 = 1$, $k = 0.01$, $p = 0.3$ (these parameters were used to fit data digitized from (Shrestha et al. 2016), as shown in Figure 3B). (B) Simulation of cell depletion as a result of activity of some hypothetical drug $C(t)$ as described by generic model (12). (C) Simulated concentration of hypothetical drug $C(t)$ that increases cell mortality according to System (12) with $k_e = 0.1$, $k_1 = 0.1$. Fifteen doses of the simulated drug $C(t)$ are given every 14 days, at a dose=1 unit; the values for drug-related parameters and schedule of administration were chosen arbitrarily for illustrative purposes.

**Discussion**

Finding a simple mechanistic ODE model to describe linear rather than exponential decrease in population size presents an unexpected challenge. Such models are nevertheless necessary to describe the dynamics of, for instance, red blood cells, which through radiolabeling experiments have been reported to decrease not exponentially but linearly over a span of approximately 120 days (Mock et al. 2011; Mock et al. 1999; Arias & Arias 2017). Using a zero-order model of population decay assumes that population size decreases independently of population density, and thus eventually predicts negative population size unless a mechanistic switch is introduced numerically when population size becomes zero. Alternatively, complex mathematical models using, for instance, PDEs (Bélair et al. 1995; Higgins & Mahadevan 2010) or integro-differential equations have been introduced to describe RBC lifespan (Shrestha et al. 2016). To our knowledge, however, until now no simple unified ODE model has been developed to describe linear population decay.

Here, we propose that linear decay can be the result of natural dynamics of a population that is heterogeneous with respect to individual death rates (if the cells had the same death rate the population would decline exponentially). We can then show that if these rates fall within a Gamma distribution, then the dynamics of the total population can be described by a sub-exponential model that can indeed predict decline pattern that is better approximated by linear rather than exponential functional form. This model can then easily fit experimentally available data on red blood cell extinction (Figure 3). Notably, it was proven in (Karev & Kareva 2016) that the underlying structure of such a model is that of an inhomogeneous population with frequency-dependent death. That is, a population where each individual (in this case a cell) is characterized by its own value of death rate parameter and where cells die proportionally to their frequency in the population (F-model) will exhibit linear decay.

F-models of population decline that predict linear-exponential decay are mirrored by F-models of population growth, which predict exponential-linear growth (Kareva & Karev 2018; Kareva & Karev 2019). In these models, subpopulations with different individual growth rates grow proportionally to their frequency in the population, and one interpretation of such frequency-dependent structure is that total population growth is proportional to access of different subpopulations to some resource, i.e., for cancer cells it can be access to blood vessels (Kareva & Karev 2018). If an F-model of extinction indeed captures the biology of RBC death, then perhaps there may exist some external factors or resources that are distributed proportionally to the subpopulations' frequencies; the absence of such resource can make individual cells more susceptible to elimination.

One can speculate about such "resource" as follows: RBC life cycle starts in the bone marrow from hematopoietic stem cell progenitors in the process called erythropoiesis. After a series of maturation steps, cells called reticulocytes enter circulation, which then mature into erythrocytes after 24-48 hours. Erythrocytes then circulate throughout the body for approximately 120 and become removed if they age or become damaged. RBC removal occurs in spleen, liver or bone marrow, where they are removed by macrophages that engulf and break them down, releasing iron stores from the cells, which, together with other factors, can stimulate further erythropoiesis (Arias & Arias 2017).

Within the framework of the proposed model, it is possible that some cells suffer uneven damage as they circulate through the body, making physical aspects of the space that is encountered during their life cycle a "death resource". Perhaps other external factors, such as exposure to phagocytic cells, may be non-uniform and thus also provide a mechanism for why

RBC death would be not density but frequency dependent. Such hypotheses remain to be evaluated. The fact that life span of RBCs may not be fixed, as is indicated by the fact that RBC lifespan is shorter in infants rather than adults (Pearson 1967), or that RBC lifespan is affected by different conditions, such as change in altitude (Risso et al. 2007) or even space flight (Alfrey et al. 1996; Rice & Alfrey 2005) could support the hypothesis of external "resources" regulating cell survival.

It is possible that more populations may be subject to linear rather than exponential decay but the lens to see and describe it was previously missing. The proposed model provides a possible mechanism that may explain such dynamics, allowing expansion of questions and hypotheses that can now be explored using mathematical modeling.

**Acknowledgements**

The authors report no external sources of funding.

**Conflicts of interest**

IK is an employee of EMD Serono, a US subsidiary of Merck, KGaA. The views expressed in this paper are the authors' personal views and do not necessarily represent the views of EMD Serono.